# Learning regularization and intensity-gradient-based fidelity for single image super resolution

Hu Liang, Shengrong Zhao *

*College of Computer Science and Technology*

*Qilu University of Technology (Shandong Academy of Sciences), 250023，Jinan, China*

*corresponding author: zhaosr2006@126.com

**Abstract:** How to extract more and useful information for single image super resolution is an imperative and difficult problem. Learning-based method is a representative method for such task. However, the results are not so stable as there may exist big difference between the training data and the test data. The regularization-based method can effectively utilize the self-information of observation. However, the degradation model used in regularization-based method just considers the degradation in intensity space. It may not reconstruct images well as the degradation reflections in various feature space are not considered. In this paper, we first study the image degradation progress, and establish degradation model both in intensity and gradient space. Thus, a comprehensive data consistency constraint is established for the reconstruction. Consequently, more useful information can be extracted from the observed data. Second, the regularization term is learned by a designed symmetric residual deep neural-network. It can search similar external information from a predefined dataset avoiding the artificial tendency. Finally, the proposed fidelity term and designed regularization term are embedded into the regularization framework. Further, an optimization method is developed based on the half-quadratic splitting method and the pseudo conjugate method. Experimental results indicated that the subjective and the objective metric corresponding to the proposed method were better than those obtained by the comparison methods.

## 1 Introduction

The conditions in real-world scenes are often far from optimal [1]. For example, taking a photograph is often restricted by the limitations of the physical resolution of the imaging devices and the imaging environments. These not only decrease the image resolution but also add blur and noise to the observation, resulting in the degradation of the image quality. In fact, the problem of "***How to increase resolution while removing the noise and blur***?" is of great interest in both theory and practice.

To overcome this difficulty, the super-resolution (SR) [2, 3] technology was proposed. According to the number of low-resolution (LR) images, the SR approach can be divided into two categories: single-frame SR method [4, 5] and multi-frame SR method [6, 7]. In this study, we addressed the problem of reconstructing an HR image from a single LR image.

There are two types of methods to solve the single-frame SR problem: regularization-based method [8-10] and deep learning-based method [11-13]. In the former method, the HR image is obtained by solving the model constructed by a fidelity term and a regularization term. The fidelity term is used to maintain the fidelity between the HR image and LR image. Existing artificial artifacts and losing high frequency detail information tend to bothering us. As far as we know, most methods preserve one of the important features (i.e., edges, textures or structures) and ignore other features, which is difficult to describe all these features in the regularization term. Moreover, limited information is a vital challenge to reconstruct an HR image with good quality, especially when the degradation is severe. Oppositely, deep learning-based method can exploit abundant information from the external training data, and it has achieved many expected results. However, it would obtain unsatisfactory results when the testing data has few similar counterparts with the external data.

In order to solve the above-mentioned problem, in this paper we combine an improved optimization method with a new designed deep learning-based method through an iterative framework. The improved optimization method is to mix the intensity and structure change in the target model and the new designed deep learning framework is a symmetric residual net. In the framework, the input is first super-resolved by optimizing the reconstruction model, which is constrained by the intensity change and structural change consistency. Then we design a convolutional neural network as external method to calculate the finally result. In the proposed algorithm, there are three major contributions which could effectively improve the reconstruction quality:

a) We design a new fidelity term which can ensure the consistency of the estimation with the observation via imposing constrains to both the intensity change and detail change for the internal SR.

b) A residual network with symmetric structure is designed to conduct the external SR. The network contains encoding and decoding layers. The encoding layer increases the number of channels and reduces its spatial dimension while the decoding layer integrates its channels into spatial details.

c) We propose a unified SR framework that integrates the external and internal information on the basis of the half quadratic splitting method.

The remainder of this paper is organized as follows. Section 2 reviews the related background briefly. Section 3 describes the proposed single-frame SR algorithm and presents the model based on the joint consistency constraint, and the CNN network. The experimental results and analyses are discussed in Section 4. And the conclusion is presented in Section 5.

## 2. Related work

### 2.1 The observation model

Let $x$ and $y$ be the HR image and LR image, respectively. In a single SR problem, the observation $y$ can be modeled as the output of a system $W$, whose input is $x$. The target of this problem is to estimate $x$ from the given data $y$ and a part or full knowledge of the system; thus, it is an inverse problem. The observation model can be written as follows:

$$y = Wx + \varepsilon \tag{1}$$

where $W$ denotes the system matrix. In general, $W = DH$, where $D$ is the down-sampling matrix, $H$ is the blurring matrix, and $\varepsilon$ denotes the noise in the observed data.

### 2.2 Regulation-based SR method

In the regularization-based method, SR can be expressed as energy minimization problems of the following type,

$$x = \arg\min_{x} L(x,y) + \lambda \Phi(x) \tag{3}$$

where, $x$ denotes the estimated HR image, $y$ denotes the observation data, $L(x, y)$ is the data fidelity term, $\Phi(x)$ is the regularization term, and $\lambda > 0$ is the regularization coefficient, which is used to balance the data fidelity term and the regularization term.

## A. Data fidelity term choice

The data fidelity term is based on the observation model and provides a restriction that the simulated LR image should be consistent with the observation [14]. Thus, the data fidelity term can ensure the consistency of the estimation $x$ with the observation $y$. The popularly used form of the fidelity term can be written as $\varphi(y - WX)$, which is used to measure the difference between the LR image and the projected estimate of the HR image. In the past few years, many fidelity terms have been proposed and they have focused on the form of $\varphi$ [15–17].

In the existing literature for the inverse problem, a large number of estimators have been designed for the data fidelity term, such as the L1, L2, Huber, and Lorentzian estimators. The L1, L2, Huber, and Lorentzian estimators can be expressed using the following expressions, respectively:

$$\varphi_{L_1}(t) = |t| \tag{4}$$

$$\varphi_{L_2}(t) = |t|^2 \tag{5}$$

$$\varphi_{Huber}(t) = \begin{cases} t^2, & t < h \\ 2h|t| - t^2, & t \geq h \end{cases} \tag{6}$$

$$\varphi_{Lorentzian}(t) = \log(1 + \frac{1}{2}\left(\frac{x}{l}\right)^2) \tag{7}$$

These fidelity terms pay attention to punishments for different level errors in the spatial domain, which cannot preserve the details effectively [18].

## B. Regularization term choice

The regularization term is used to model the prior knowledge and penalize the plausible solutions that do not satisfy the assumed properties, working as a constraint on the estimated HR image. Most regularization-based approaches focus on designing the regularization term, which has a tremendous influence on the SR performance. Thus, many effective and commonly used priors for the SR problem have been proposed in the literature, such as the total variation (TV)-type prior [19, 20], sparsity prior [21], nonlocal prior [22], adaptive prior [23], steering kernel regression prior [24], gradient profile prior [25], and combination prior [26].

The regularization term is usually artificially designed; thus, the constraint imposed by the prior knowledge owns an artificial bias. For example, the Tikhonov prior model is based on the L2 norm, and it can penalize large gradient, which often blurs the image edges. The TV-type prior model is based on the L1 norm, which can preserve the large gradients and lead to over-smoothing.

**2.3 Deep learning-based SR method**

With the seminal work, super-resolution convolutional neural network (SRCNN) [27], a convolutional network model is introduced to image SR. And the deep learning has been actively applied to the SR field. On the basis of SRCNN, Kim [28] increased the depth of network to 20 layers and used the residual learning (VDSR), achieving better convergence. For better learning accuracy, He [29] used residual dense block based SR method, i.e., ResNet, to extract local features. In [30], the deep recursive residual network (DRRN) is proposed. This model contains 52 convolution layers, and performs residual learning at both local and global levels. Lai [31] proposed a coarse-to-fine Laplacian pyramid framework, i.e., the Laplacian pyramid SR network, for image super-resolution model. In [32], a SR method based on geometric similarity is proposed, which focuses on digging the potential information in the image itself. In [33], an edge guided recurrent residual network is proposed, thus, this method has edge-preserving capability, which could make the reconstructed image conform to the human perception. In [34], the generative adversarial network is applied to image SR, which could generate the image details effectively. In order to capture larger scale contextual image information, Zhang [35] used dilated convolutions for image SR, whose convolutional layer contains both standard and dilated convolutions. However, larger receptive fields increase the network depths, which may cause degradation. Zheng [36] proposed a joint residual pyramid model for reconstructing HR images, which combined residual blocks and linear interpolation layers, and applied them into a convolutional neural pyramid. In recent years, the learning based methods are trying to be combined with other methods. In [37], a hybrid SR method is proposed, which exploited the external and internal SR methods, which are conducted by a CNN model and self-similarity, respectively. In [38] and [39], the learned CNN denoisers are introduced into the model-based optimization methods as a modular part to solve the restoration problem. This strategy can take advantage the flexiblility of model-based optimization methods and the rapidness of CNN-based methods.

## 3. Proposed algorithm

### 3.1 Constraint for data consistency

In the regularization method, the data fidelity term is usually given by

$$F_1(x, y) = \|y - Wx\|_p^p \tag{8}$$

Corresponding to the commonly used degradation model, Eq. (8) is widely applied in image SR to describe the consistency between $y$ and $Wx$. For a specific image, it is expected to reconstruct an HR image in terms of both edges and textures. Eq.(8) only reflects the consistency in intensity which cannot preserve the details information in edge and texture properly. In the gradient fields, there still exists plenty of difference which cannot be reflected in intensity fields (seen in Figure 1).

Figure 1 shows the error images in different cases. (a) is the error calculated by $y$-$Wx$, (b) is the error obtained by $\nabla y - \nabla(Wx)$, where $\nabla$ is the gradient operator, (c) is the error obtained by $\Delta y - \Delta(Wx)$, where $\Delta$ is the Laplace operator. From Figure 1, we can see that difference exists in all cases. On one hand, the error obtained by observation $y$ and estimated image $Wx$ is concentrated on the edges or structures. On the other hand, the error obtained under $\nabla$ and $\Delta$ is focused on the flat region and texture. The three models have different concerns and their roles are hardly inter-changeable. Hence, in this work, in addition to the consistency constraint on image intensity, the consistency constraint on data gradients is also considered for SR reconstruction.

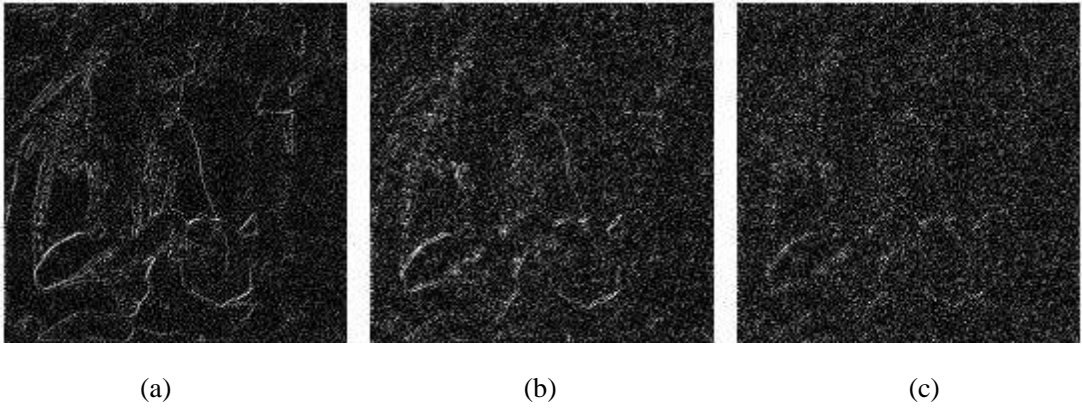

| (a) | (b) | (c) |

Figure 1. The error images. (a) is the error calculated by *y-Wx*, (b) is the error obtained by $\nabla y - \nabla(Wx)$, (c) is the error obtained by $\Delta y - \Delta(Wx)$.

Usually, the large gradients are used to depict the edges, while the small gradients are used to denote the image texture. The first-order gradient is used to preserve the edges and the second-order gradient is applied to enhance texture details. Thus, for SR framework, we establish a new fidelity term as follows,

$$\Psi(x, y) = \|\psi(y) - \psi(Wx)\|_p^p \quad (9)$$

where $\psi(\square)$ denotes the feature extraction function, such as the gradient operator, or the Laplacian operator.

### 3.2 SR based on the new fidelity term

In this paper, the data fidelity term for SR is proposed as follows:

$$F(x, y) = \sum_i \lambda_i \|\psi_i(y) - \psi_i(Wx)\|_p^p \quad (10)$$

where, $\psi_i$ denotes feature extraction function, and $\lambda_i$ are positive parameters that can control the relative importance of all the terms.

Based on Eq.(10), the corresponding objective function can be written as follows,

$$\hat{x} = \arg\min_x \ F(x,y) + \lambda \Phi(x)$$

With the half quadratic splitting (HQS) method, by introducing an intermediate variable *z*, the optimization problem was obtained as follows on the basis of the proposed data fidelity term,

$$\hat{x} = \arg\min_{x,z} F(x,y) + \lambda\Phi(z) + \gamma\|z-x\|_2^2 \tag{11}$$

and then, Eq. (11) was minimized by optimizing the following two sub-problems:

$$\begin{cases} \hat{x} = \arg\min_{x} F(\mathrm{x,y}) + \gamma\|z-x\|_2^2 \\ \hat{z} = \arg\min_{z} \lambda\Psi(z) + \gamma\|z-x\|_2^2 \end{cases} \tag{12}$$

Thus, the optimal $x$ and $z$ were obtained via the following iterative scheme:

$$\begin{cases} x_{k+1} = \arg\min_{x} F(\mathrm{x,y}) + \gamma\|z_k-x\|_2^2 & (a) \\ z_{k+1} = \arg\min_{z} \lambda\Psi(z) + \gamma\|z-x_{k+1}\|_2^2 & (b) \end{cases} \tag{13}$$

### 3.3 Optimization

**A. Solving the x-sub-problem**

By setting the first derivative of the cost function of x-sub-problem to zero, we obtained the following:

$$(W^T W + W^T \nabla^T \nabla W + W^T \Delta^T \Delta W + \gamma I)x_{k+1} = W^T y + W^T \nabla^T \nabla y + W^T \Delta^T \Delta y + \gamma z_k \tag{14}$$

Eq. (14) has the form of $Ax = b$ and it is a large, sparse equation. Thus, the preconditioned conjugate gradients method was applied to solve Eq. (14) iteratively, which could solve the problem efficiently.

**B. Solving the z-sub-problem**

Obviously, Eq. (13b) can be seen as a de-noising problem in which $z$ is the variable to be solved. In general, the solution is to establish the regularization term $\Phi(z)$. Recent study [38-40] shows that the regularizer can be replaced by the implicit prior learned from the neural network.

In this paper, the denoising problem Eq.(13b) was achieved by using a CNN-based residual network, as shown in Figure 2. The network is a symmetric, the left part is based on convolution (encoder) and the right part is on the basis of deconvolution (decoder). In our work, we set 5 convolution blocks and 5 deconvolution blocks. Each block contains a convolution (encoder) or deconvolution (decoder), Bnorm (i.e., the batch normalization) and ReLU (i.e., the rectified linear units) operation. The number of channels is 256, 128, 64, 32 and 16 respectively in each convolution layer. The purpose of residual links can be understood in the context of residual–nets: they allow to constructing the solution using both coarse and processed data.

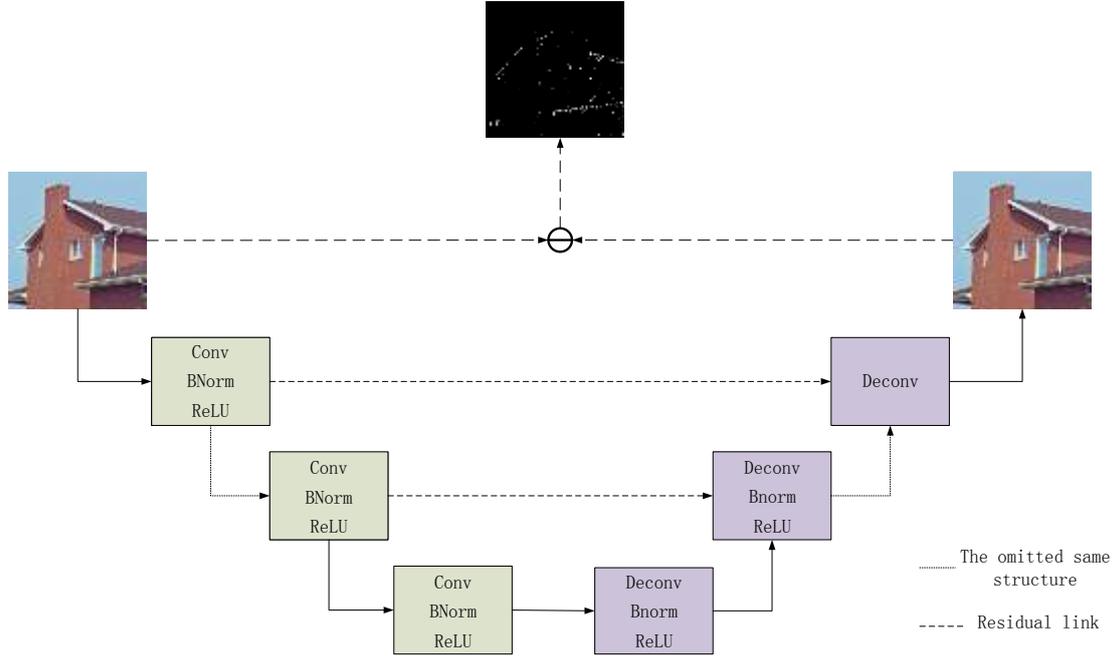

Figure 2. The applied CNN network.

**C. Insight of choosing the network for denoising**

Defining a good regularization term is crucial and difficult in image SR process. The constraint imposed by the artificially designed prior model owns artifacts. Traditionally, the regularization term is usually artificially designed. Hence they are widely adopted in model-based optimization methods to solve the inverse problems. The popular image prior models include the TV-type prior, sparsity prior, nonlocal prior, steering kernel regression prior, and gradient profile prior [25], etc. However, the constraint imposed by these prior knowledge owns artificial bias, thus the priors have their respective drawbacks. For example, the Tikhonov prior model often blurs the image edges because it can penalize a large gradient. The TV-type prior model often leads to over-smoothing. Thus, efficiently image prior is highly demanded.

With this network, the denoising results, which were also the solution of Eq. (13b), were obtained. There was no need to know the explicit form of the regularization term when this residual network was used. And, a deep CNN network can exhibit powerful prior modeling capacity. Moreover, due to the symmetric structure, the encoding layer increases the number of channels and reduces its spatial dimension. On the contrary, the decoding layer integrates its channels into spatial details.

## 4. Numerical implementation and experimental results

In this section, we will discuss the several experiments conducted to evaluate the effectiveness of the proposed method. The programming environment to carry out the experiments was MATLAB R2015b on an Intel(R) Core (TM) i7-6700 PC.

Many methods published in recent years were selected for comparison, namely the bicubic interpolation method, SRCNN [27], VDSR [28], DRRN [30], the single-image SR method proposed in [38] (called SISR for short), A+ [41], SR using collaborative representation cascade (SRCRC) [42], SR using collaborative representation and non-local self-similarity (SRCRNS) [43], SR via multiple mixture prior models (SRMMPM) [44]. All the results of the compared methods were generated by using the respective authors' codes, and the parameters were set according to their papers or tuned to generate the best results.

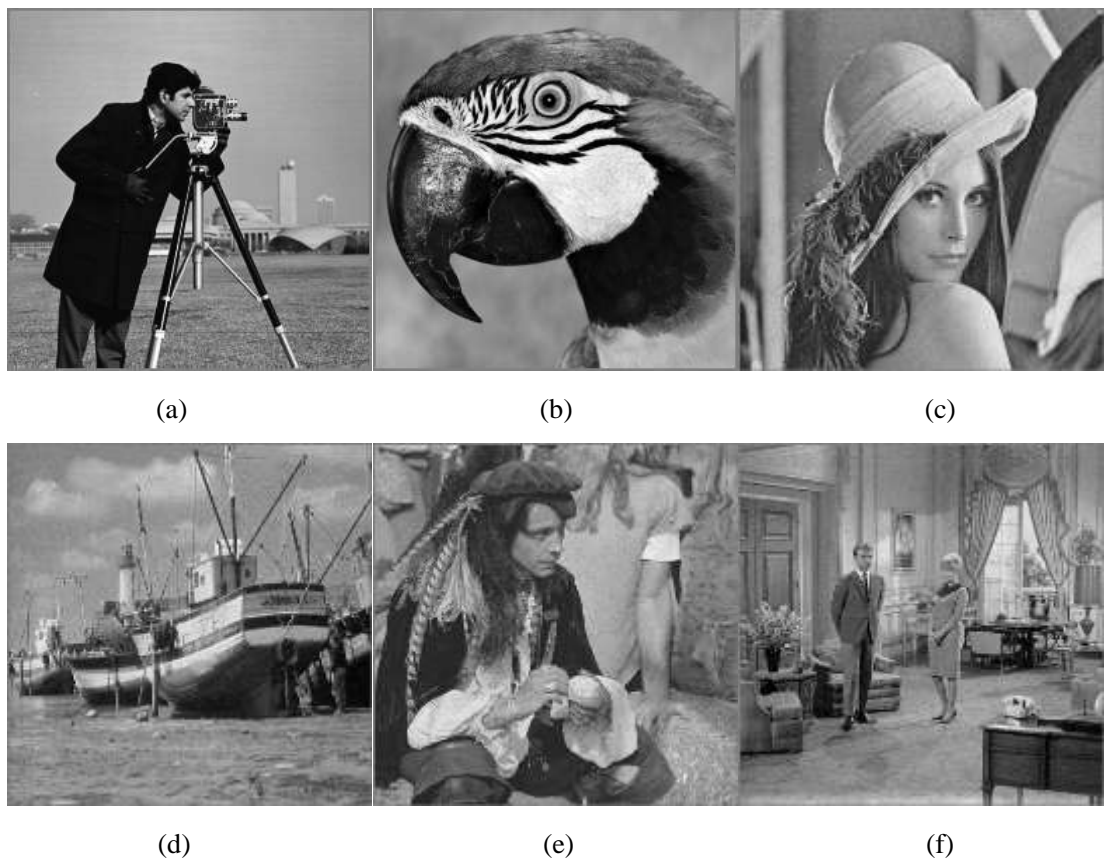

Figure 3. Test images. **a** Cameraman, **b** Parrots, **c** Lenna, **d** Boat, **e** Man, **f** Couple.

In order to evaluate the SR performance obtained by different methods, both subjective methods and objective methods were used to measure and compare the quality of the estimated HR images. The subjective assessment indices included the reconstructed images and the error images. The error images showed the difference between the reconstructed images and the original image. In the error images, the larger the number of white points was, the larger the number of reconstruction errors was. Moreover, the two objective assessment indices of the peak signal-to-noise ratio (PSNR) and structural similarity (SSIM) were used. PSNR is the most commonly and widely used objective assessment index. However, in many cases, PSNR values cannot be completely consistent with the visual quality. Therefore, we also used SSIM to evaluate the reconstructions objectively. SSIM and PSNR provided indications of image quality on the basis of the known characteristics of the visual system of humans.

**4.1 Experimental setup**

Six test images (presented in Figure 3) were used to demonstrate the improvement of the proposed method. In the experiments, the LR image was generated from the original image by blurring and down-sampling. First, the original image was produced by blurring the original image with a blur kernel. In the experiments, two types of kernels were used: the Gaussian kernel and the average kernel. Two types of Gaussian kernels were used: $3\times3$ with variance $\sigma_{blur}=1$, $5\times5$ with $\sigma_{blur}=2$, and two types of average kernels were used: $3\times3$ and $5\times5$. Then, the LR image was obtained by discarding every other row and column of the blurred image. That is, the LR image was down-sampled by a factor of 2. Finally, additive Gaussian white noises with variance $\sigma_{noise}=1$, $\sigma_{noise}=3$, and $\sigma_{noise}=5$ were added to the obtained images with low quality.

For data training, 5,000 images were collected from BSD, ImageNet database, and Waterloo Exploration database. The images were cropped into small patches of size $35\times35$, and 2,500,000 patches were selected for training. Moreover, additive

Gaussian noise was added to the clean patches during training so as to obtain the corresponding noisy patches. The training procedure was achieved by using the following cost function:

$$\hat{\theta} = \arg\min_{\theta} \frac{1}{2N} \sum_{i=1}^{N} \| f(Y_i, \theta) - (Y_i - X_i) \|_2^2 \quad (15)$$

where $f(\cdot)$ is the denoising mapping learned by the CNN network, $\theta$ is the CNN weight, and $\{Y_i, X_i\}_{i=1}^{N}$ represents $N$ noisy-clean patch pairs.

**4.2 Benefits of joint fidelity term**

To validate the effectiveness of the combined data fidelity term, a set of experiments using two different versions of the proposed SR method were conducted. In the first version, $i=1$ and $\psi_1(t)=t$, and in the second version, $i=3$ and $\psi_1(t)=t, \psi_2(t)=\nabla t, \psi_3(t)=\Delta t$. Here, the proposed two versions are called Pro.v1 and Pro.v2, respectively.

Table 1 illustrates the PSNR and SSIM values of the two different versions on the six test images shown in Figure 2, respectively, under $3\times 3$ Gaussian blur with $\sigma_{blur}=1$ and/or $3\times 3$ average blur, and additive Gaussian white noises with $\sigma_{noise}=1$. An analysis of the results presented in Table 1 revealed that Pro.v2 achieved better performance than Pro.v1, which demonstrated that the combination of the L2-based fidelity term, gradient-based fidelity term, and the Laplacian-based fidelity term was helpful in enhancing the quality of the reconstructed HR images. To compare the visual quality of the reconstructions produced by the two versions, the estimated error results of the image Man and the image Couple are presented in Tables 2 and 3. We observed that the error images corresponding to Pro.v2 contained less white points than those corresponding to Pro.v1. Thus, we concluded that the results of Pro.v2 were more reliable. The experiment results presented in this subsection indicated that the combination of the three terms was helpful in improving the SR performance. Moreover, the statistical data of the error images are provided in Tables 2 and 3. The three classical statistical indicators were the absolute maximum error (max), the mean value (mean) of the absolute error, and the variance (var) of the absolute error between the original image and the reconstructed images. Smaller data presented better results. From Tables 2 and 3, we inferred that Pro.v2 gained a larger increase in statistical results.

Table 1. PSNR (dB) /SSIM comparison of Pro.c1 and Pro.v2.

| Figure | Blur/noise | Method | | Figure | Blur/noise | Method | |
| --- | --- | --- | --- | --- | --- | --- | --- |
| | | Pro.v1 | Pro.v2 | | | Pro.v1 | Pro.v2 |
| Cameraman | GB/$\sigma_{noise}$=1 | 26.22/0.8429 | 27.43/0.8698 | Boat | GB/$\sigma_{noise}$=1 | 29.58/0.8286 | 31.09/0.8604 |
| | GB/$\sigma_{noise}$=3 | 26.15/0.8377 | 27.28/0.8610 | | GB/$\sigma_{noise}$=3 | 29.45/0.8233 | 30.77/0.8503 |
| | GB/$\sigma_{noise}$=5 | 26.01/0.8214 | 27.92/0.8383 | | GB/$\sigma_{noise}$=5 | 29.17/0.8059 | 30.15/0.8204 |
| | AB/$\sigma_{noise}$=1 | 25.83/0.8320 | 27.29/0.8624 | | AB/$\sigma_{noise}$=1 | 29.15/0.8170 | 30.61/0.8490 |
| | AB/$\sigma_{noise}$=3 | 25.79/0.8271 | 26.94/0.8507 | | AB/$\sigma_{noise}$=3 | 29.05/0.8124 | 30.00/0.8341 |
| | AB/$\sigma_{noise}$=5 | 25.66/0.8102 | 27.26/0.8295 | | AB/$\sigma_{noise}$=5 | 28.76/0.7267 | 29.78/0.8088 |
| Parrots | GB/$\sigma_{noise}$=1 | 26.69/0.8670 | 28.63/0.8941 | Man | GB/$\sigma_{noise}$=1 | 29.89/0.8469 | 31.34/0.8811 |
| | GB/$\sigma_{noise}$=3 | 26.67/0.8638 | 28.20/0.8852 | | GB/$\sigma_{noise}$=3 | 29.77/0.8424 | 30.97/0.8712 |
| | GB/$\sigma_{noise}$=5 | 26.50/0.8494 | 28.42/0.8568 | | GB/$\sigma_{noise}$=5 | 29.43/0.8278 | 30.22/0.8356 |
| | AB/$\sigma_{noise}$=1 | 26.22/0.8572 | 28.11/0.8844 | | AB/$\sigma_{noise}$=1 | 29.47/0.8345 | 30.81/0.8673 |
| | AB/$\sigma_{noise}$=3 | 26.20/0.8538 | 27.78/0.8755 | | AB/$\sigma_{noise}$=3 | 29.38/0.8304 | 30.54/0.8585 |
| | AB/$\sigma_{noise}$=5 | 26.07/0.8412 | 27.89/0.8489 | | AB/$\sigma_{noise}$=5 | 29.06/0.8158 | 29.92/0.8256 |
| Lenna | GB/$\sigma_{noise}$=1 | 32.65/0.8902 | 34.49/0.9079 | Couple | GB/$\sigma_{noise}$=1 | 29.12/0.8251 | 30.36/0.8589 |
| | GB/$\sigma_{noise}$=3 | 32.49/0.8854 | 34.07/0.8991 | | GB/$\sigma_{noise}$=3 | 29.04/0.8216 | 30.63/0.8584 |
| | GB/$\sigma_{noise}$=5 | 32.03/0.8697 | 32.86/0.8658 | | GB/$\sigma_{noise}$=5 | 28.79/0.8099 | 29.83/0.8294 |
| | AB/$\sigma_{noise}$=1 | 32.21/0.8846 | 33.92/0.9016 | | AB/$\sigma_{noise}$=1 | 28.71/0.8107 | 29.94/0.8455 |
| | AB/$\sigma_{noise}$=3 | 32.00/0.8789 | 33.56/0.8930 | | AB/$\sigma_{noise}$=3 | 28.64/0.8078 | 29.75/0.8380 |
| | AB/$\sigma_{noise}$=5 | 31.61/0.8635 | 32.57/0.8612 | | AB/$\sigma_{noise}$=5 | 28.42/0.7958 | 29.46/0.8167 |

Table 2. The comparison of estimation error corresponding to image Man by Pro.v1 and Pro.v2 with various attacks. max means the maximum error, mean means mean error and var means the variance of the error.

| FIGURE | BLUR/ NOISE | Method | Error | std | | |
|---|---|---|---|---|---|---|
| | | | | max | mean | var |
| 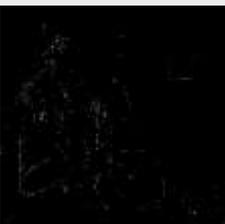 "MAN" | 3*3 Gaussian Blur/ $\sigma_{noise}=1$ | Pro.v1 | 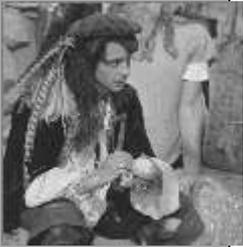 | 129 | 5.20 | 39.55 |
| | | Pro.v2 | 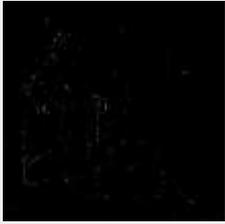 | 96 | 4.46 | 27.88 |
| | 3*3 Gaussian Blur/ $\sigma_{noise}=1$ | Pro.v1 | 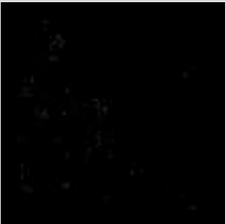 | 127 | 5.34 | 40.07 |
| | | Pro.v2 | 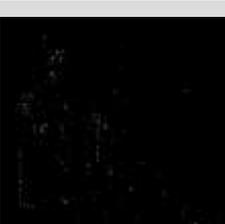 | 93 | 4.75 | 29.41 |
| | 3*3 Gaussian Blur/ $\sigma_{noise}=1$ | Pro.v1 | 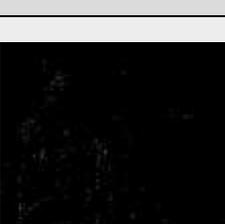 | 123 | 5.71 | 41.46 |
| | | Pro.v2 | 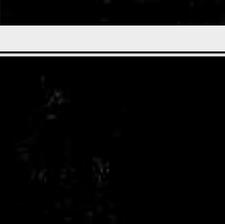 | 89 | 5.50 | 31.61 |

Table 3. The comparison of estimation error corresponding to image Couple by Pro.v1 and Pro.v2 with various attacks. max means the maximum error, mean means mean error and var means the variance of the error.

| FIGURE | BLUR/NOISE | method | error | std | | |
|---|---|---|---|---|---|---|
| | | | | max | mean | var |
| "COUPLE" 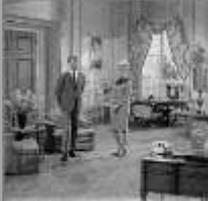 | 3*3 Gaussian Blur/ $\sigma_{noise}=1$ | Pro.v1 | 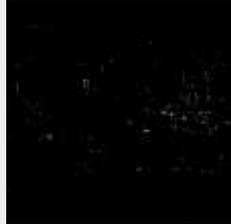 | 94 | 5.95 | 44.21 |
| | | Pro.v2 | 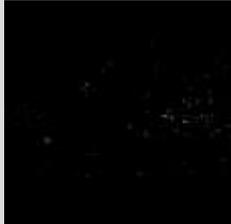 | 70 | 5.22 | 32.57 |
| | 3*3 Gaussian Blur/ $\sigma_{noise}=1$ | Pro.v1 | 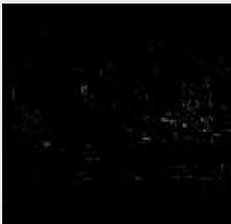 | 92 | 6.04 | 44.64 |
| | | Pro.v2 | 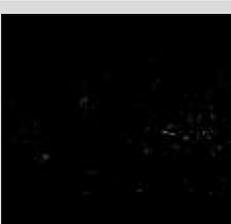 | 67 | 5.19 | 29.24 |
| | 3*3 Gaussian Blur/ $\sigma_{noise}=1$ | Pro.v1 | 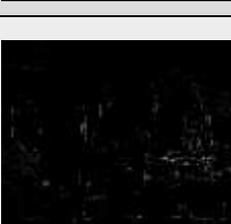 | 92 | 6.33 | 45.88 |
| | | Pro.v2 | 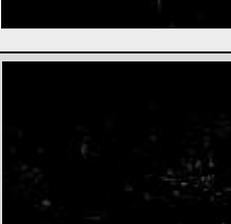 | 66 | 5.62 | 32.25 |

### 4.3 Experiments on simulated data with down-sampling factor 2

In this subsection, the effectiveness of the proposed method (i.e., $i=3$ and $\psi_1(t)=t, \psi_2(t)=\nabla t, \psi_1(t)=\Delta t$) was examined under the case of down-sampling factor equaling to 2. We compared the proposed method with the bicubic interpolation method, the SRCNN method, the DRRN method, and the SISR method. The test $256 \times 256$ and $512 \times 512$ images included Cameraman, Peppers, Lena, Boat, Man, and Couple shown in Figure 3. The restored Cameraman, Peppers, Lena, Boat, Man, and Couple images obtained by using different SR methods are separately listed in Figures 4 and 5. Figures 4a and 5a show the original images of Cameraman, Peppers, Lena, Boat, Man, and Couple. Figures 4b and 5b show the results obtained by using the bicubic interpolation method. Figures 4c and 5c show the results obtained by using the SRCNN method. Figures 4d and 5d show the results obtained by using the DRNN method. Figures 4e and 5e show the results obtained by using the SISR method. Figures 4f and 5f show the results obtained by using the proposed method.

For a detailed comparison, we present the estimation error corresponding to Figures 4 and 5, respectively. Note that in the error images, the larger the number of white points was, the larger was the number of reconstruction errors.

From Figures 4 and 5, we inferred that the proposed method produced cleaner results, and it had finer reconstruction details than the compared methods, which was reflected by the comparison of the error images presented in Figures 4 and 5 (g) – (l). In general, among all the error images, the error images corresponding to the SRBS&GI method contained the least number of white points, which demonstrated that the reconstructions obtained by the proposed method were closer to the original images. In our method, the constraints in the gradient domain were included; thus, it performed better in terms of detail preservation. The visual comparison of different SR methods showed that the proposed method could preserve the detail information and remove the artifacts effectively.

Besides the subjective comparison, we conducted an objective comparison of the proposed method and the compared methods. A quantitative evaluation is performed

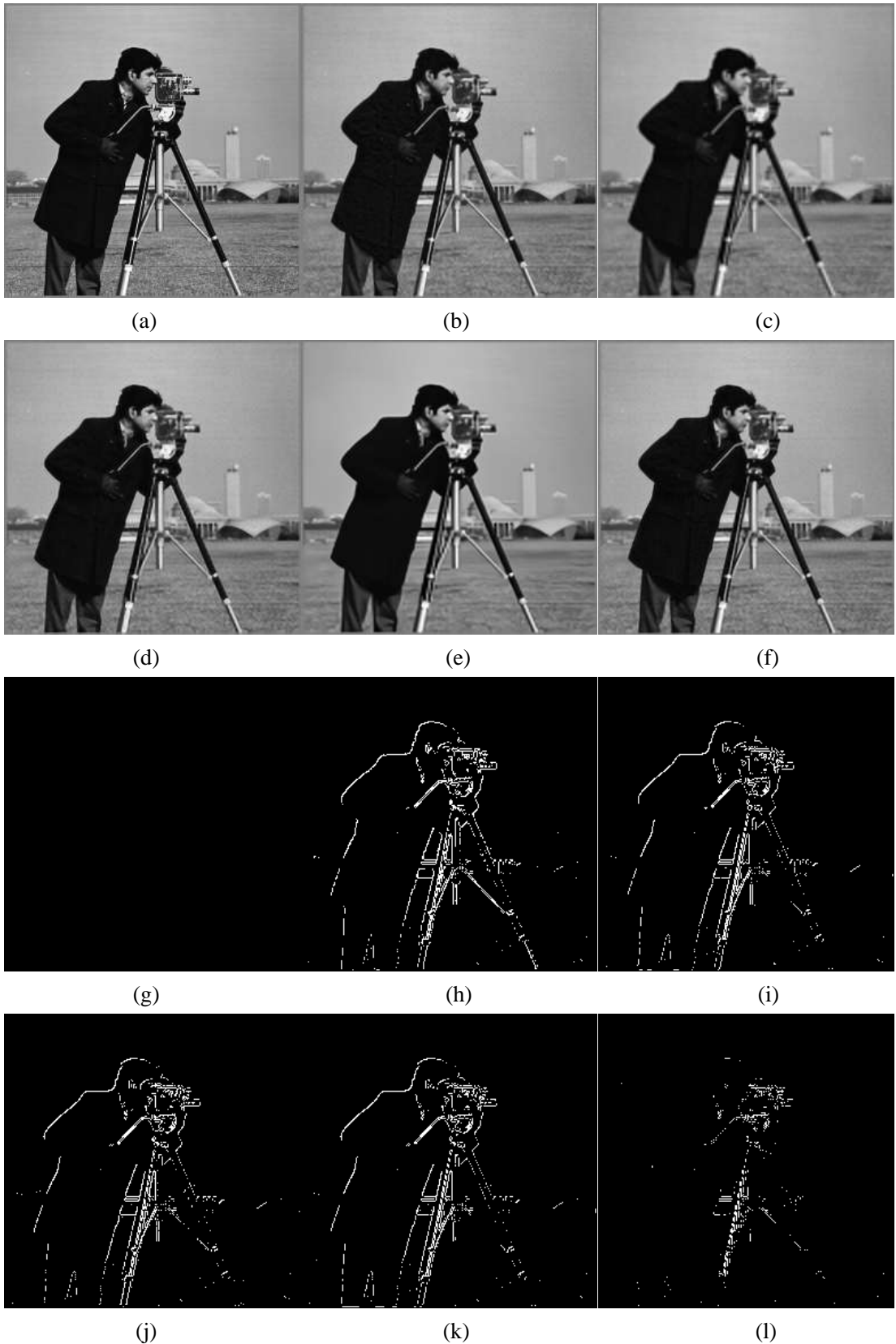

Figure 4. Reconstructed results of comparison of different methods on Cameraman with a scale factor 2, 5×5 Gaussian kernel blur of 1 and Gaussian noise of 1. (a) Original image; (b) Bicubic; (c) SRCNN; (d) DRNN; (e) SISR; (f) Proposed. Visualization of estimation error corresponding to (a) to (f) are presented in (g) to (l).

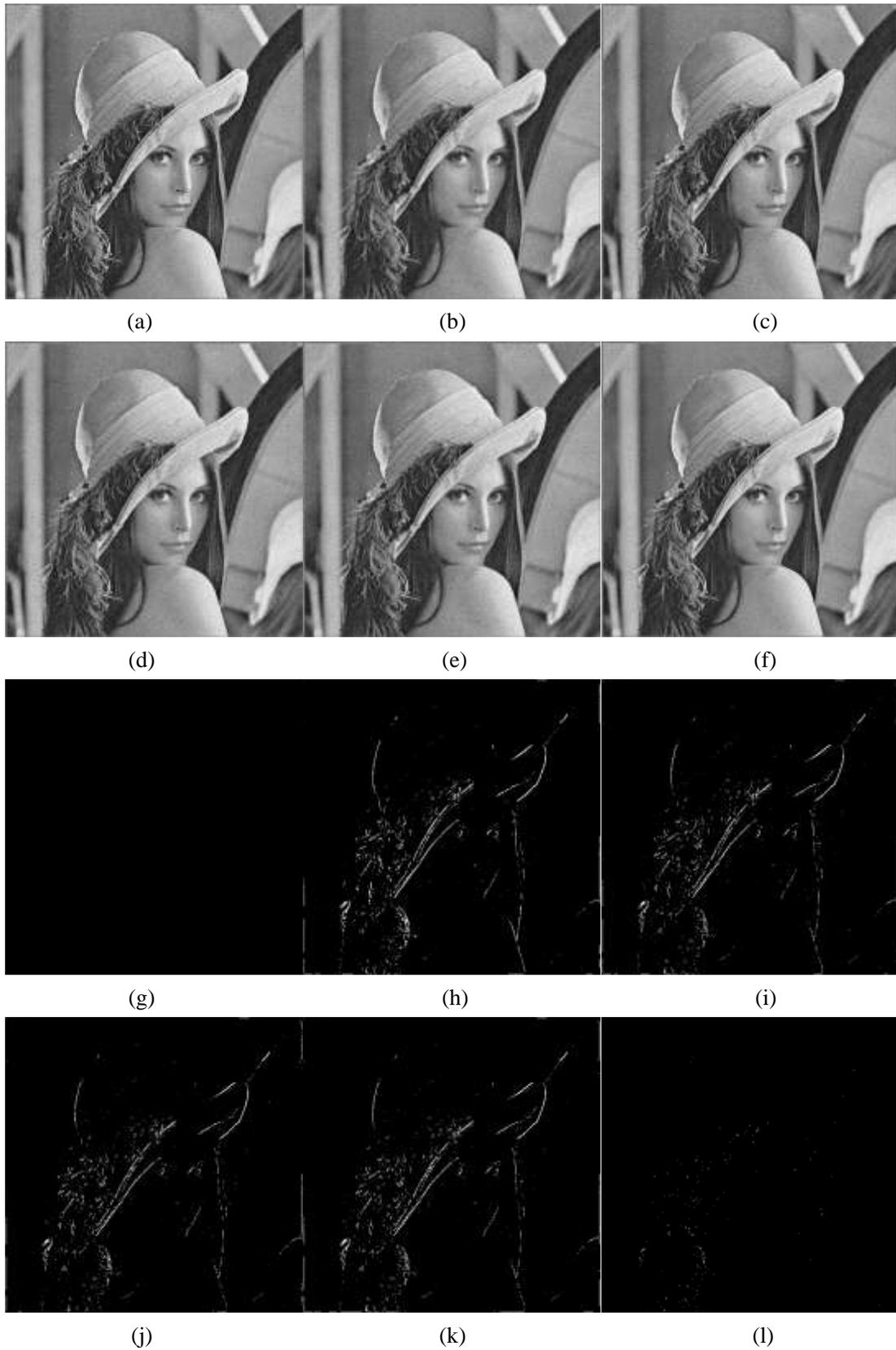

Figure 5. Reconstructed results of comparison of different methods on Lenna with a scale factor 2, 3×3 Average kernel blur and Gaussian noise of 3. (a) Original image; (b) Bicubic; (c) SRCNN; (d) DRNN; (e) SISR; (f) Proposed. Visualization of estimation error corresponding to (a) to (f) are presented in (g) to (l).

Table 4. PSNR/SSIM comparisons of our proposed method with state-of-the-art methods under 3*3 and 5*5 Gaussian kernel and noise level σnoise=1, σnoise=3, and σnoise=5.

| Figure | Gaussian kernel | Method | | | | |
|---|---|---|---|---|---|---|
| | | Bicubic | SISR | SRCNN | DRRN | Proposed |
| Cameraman | 3*3/$\sigma_{noise}$=1 | 24.32/0.8116 | 24.63/0.8331 | 24.53/0.8288 | 24.51/0.8247 | **27.43/0.8698** |
| | 3*3/$\sigma_{noise}$=3 | 24.22/0.7798 | 24.36/0.7500 | 24.30/0.7611 | 24.24/0.7461 | **27.28/0.8610** |
| | 3*3/$\sigma_{noise}$=5 | 24.04/0.7226 | 23.85/0.6314 | 23.83/0.6551 | 23.72/0.6336 | **27.92/0.8383** |
| | 5*5/$\sigma_{noise}$=1 | 23.15/0.7499 | 23.31/0.7590 | 23.41/0.7589 | 23.35/0.7514 | **24.33/0.7795** |
| | 5*5/$\sigma_{noise}$=3 | 23.08/0.7176 | 23.12/0.6744 | 23.24/0.6905 | 23.18/0.6718 | **24.32/0.7717** |
| | 5*5/$\sigma_{noise}$=5 | 22.92/0.6643 | 22.72/0.5648 | 22.87/0.5920 | 23.80/0.7441 | **24.18/0.7459** |
| Parrots | 3*3/$\sigma_{noise}$=1 | 24.13/0.8455 | 24.45/0.8579 | 24.33/0.8570 | 24.43/0.8548 | **28.63/0.8941** |
| | 3*3/$\sigma_{noise}$=3 | 24.05/0.8185 | 24.20/0.7848 | 24.13/0.7945 | 24.16/0.7839 | **28.20/0.8852** |
| | 3*3/$\sigma_{noise}$=5 | 23.86/0.7731 | 23.68/0.6899 | 23.71/0.7074 | 23.69/0.6907 | **28.42/0.8568** |
| | 5*5/$\sigma_{noise}$=1 | 22.74/0.7873 | 23.00/0.7918 | 23.06/0.7959 | 23.01/0.7910 | **24.36/0.8101** |
| | 5*5/$\sigma_{noise}$=3 | 22.69/0.7601 | 22.81/0.7192 | 22.90/0.7342 | 22.83/0.7216 | **24.38/0.8060** |
| | 5*5/$\sigma_{noise}$=5 | 22.56/0.7155 | 22.47/0.6250 | 22.58/0.6466 | 22.49/0.6277 | **24.29/0.7823** |
| Lenna | 3*3/$\sigma_{noise}$=1 | 29.85/0.8733 | 30.23/0.8786 | 30.10/0.8786 | 30.13/0.8788 | **34.49/0.9079** |
| | 3*3/$\sigma_{noise}$=3 | 29.53/0.8423 | 29.31/0.7939 | 29.29/0.8028 | 29.25/0.7971 | **34.07/0.8991** |
| | 3*3/$\sigma_{noise}$=5 | 28.93/0.7891 | 27.88/0.6809 | 27.97/0.6960 | 27.86/0.6860 | **32.86/0.8658** |
| | 5*5/$\sigma_{noise}$=1 | 28.51/0.8327 | 28.90/0.8353 | 28.77/0.8360 | 27.78/0.8357 | **30.32/0.8525** |
| | 5*5/$\sigma_{noise}$=3 | 28.26/0.8022 | 28.19/0.7517 | 28.15/0.7615 | 28.11/0.7550 | **30.20/0.8448** |
| | 5*5/$\sigma_{noise}$=5 | 27.85/0.7500 | 27.09/0.6386 | 27.16/0.6549 | 27.06/0.6441 | **29.84/0.8180** |
| Boat | 3*3/$\sigma_{noise}$=1 | 27.52/0.7987 | 27.87/0.8189 | 27.74/0.8143 | 27.72/0.8162 | **31.09/0.8604** |
| | 3*3/$\sigma_{noise}$=3 | 27.33/0.7761 | 27.32/0.7564 | 27.24/0.7594 | 27.18/0.7562 | **30.77/0.8503** |
| | 3*3/$\sigma_{noise}$=5 | 26.97/0.7380 | 26.37/0.6684 | 26.37/0.6772 | 26.26/0.6702 | **30.15/0.8204** |
| | 5*5/$\sigma_{noise}$=1 | 26.18/0.7305 | 26.51/0.7433 | 26.49/0.7432 | 26.47/0.7426 | **27.41/0.7576** |
| | 5*5/$\sigma_{noise}$=3 | 26.03/0.7079 | 26.08/0.6813 | 26.12/0.6886 | 26.07/0.6835 | **27.35/0.7505** |
| | 5*5/$\sigma_{noise}$=5 | 25.77/0.6707 | 25.37/0.5965 | 25.46/0.6092 | 25.39/0.6004 | **27.12/0.7285** |
| Man | 3*3/$\sigma_{noise}$=1 | 28.01/0.8239 | 28.37/0.8448 | 28.36/0.8430 | 28.39/0.8456 | **31.34/0.8811** |
| | 3*3/$\sigma_{noise}$=3 | 27.80/0.8011 | 27.75/0.7806 | 27.79/0.7861 | 27.75/0.7833 | **30.97/0.8712** |
| | 3*3/$\sigma_{noise}$=5 | 27.39/0.7602 | 26.69/0.6882 | 26.81/0.6997 | 26.73/0.6929 | **30.22/0.8356** |
| | 5*5/$\sigma_{noise}$=1 | 26.73/0.7551 | 27.00/0.7670 | 26.99/0.7671 | 26.98/0.7661 | **27.90/0.7769** |
| | 5*5/$\sigma_{noise}$=3 | 26.58/0.7333 | 26.56/0.7049 | 26.60/0.7122 | 26.56/0.7069 | **27.84/0.7717** |
| | 5*5/$\sigma_{noise}$=5 | 26.27/0.6951 | 25.73/0.6156 | 25.83/0.6281 | 25.75/0.6197 | **27.56/0.7477** |
| Couple | 3*3/$\sigma_{noise}$=1 | 27.25/0.7896 | 27.72/0.8182 | 27.56/0.8112 | 27.66/0.8176 | **30.36/0.8589** |
| | 3*3/$\sigma_{noise}$=3 | 27.08/0.7701 | 27.17/0.7628 | 27.08/.7614 | 27.13/0.7641 | **30.63/0.8584** |
| | 3*3/$\sigma_{noise}$=5 | 26.75/0.7362 | 26.28/0.6830 | 26.27/0.6878 | 26.27/0.6862 | **29.83/0.8294** |
| | 5*5/$\sigma_{noise}$=1 | 25.90/0.7083 | 26.25/0.7255 | 26.22/0.7256 | 26.19/0.7237 | **26.98/0.7374** |
| | 5*5/$\sigma_{noise}$=3 | 25.78/0.6897 | 25.87/0.6707 | 25.88/0.6770 | 25.84/0.6725 | **26.94/0.7334** |
| | 5*5/$\sigma_{noise}$=5 | 25.52/0.6548 | 25.16/0.5910 | 25.23/0.6017 | 25.16/0.7077 | **26.73/0.7148** |

Table 5. PSNR/SSIM comparisons of our proposed method with state-of-the-art methods under 3*3 and 5*5 Average kernel and noise level $\sigma_{noise}=1$, $\sigma_{noise}=3$, and $\sigma_{noise}=5$.

| Figure | Average kernel | method | | | | |
|---|---|---|---|---|---|---|
| | | Bicubic | SISR | SRCNN | DRRN | Proposed |
| Cameraman | 3*3/$\sigma_{noise}$=1 | 24.13/0.8024 | 24.64/0.8284 | 24.52/0.8232 | 24.54/0.8183 | **27.29/0.8624** |
| | 3*3/$\sigma_{noise}$=3 | 24.04/0.7703 | 24.38/0.7413 | 24.30/0.7521 | 24.29/0.7373 | **26.94/0.8507** |
| | 3*3/$\sigma_{noise}$=5 | 23.87/0.7163 | 23.84/0.6308 | 23.86/0.6539 | 23.76/0.6319 | **27.26/0.8295** |
| | 5*5/$\sigma_{noise}$=1 | 22.74/0.7277 | 22.67/0.7258 | 22.79/0.7275 | 22.70/0.7187 | **23.48/0.7474** |
| | 5*5/$\sigma_{noise}$=3 | 22.68/0.6974 | 22.53/0.6450 | 22.66/0.6620 | 23.29/0.7359 | **23.47/0.7399** |
| | 5*5/$\sigma_{noise}$=5 | 22.54/0.6457 | 22.18/0.5382 | 22.32/0.5668 | 22.22/0.5412 | **23.34/0.7157** |
| Parrots | 3*3/$\sigma_{noise}$=1 | 23.92/0.8376 | 24.40/0.8533 | 24.34/0.8530 | 24.38/0.8506 | **28.11/0.8844** |
| | 3*3/$\sigma_{noise}$=3 | 23.84/0.8100 | 24.14/0.7788 | 24.10/0.7898 | 24.11/0.7788 | **27.78/0.8755** |
| | 3*3/$\sigma_{noise}$=5 | 23.67/0.7644 | 23.65/0.6825 | 23.66/0.7001 | 23.62/0.6836 | **27.89/0.8489** |
| | 5*5/$\sigma_{noise}$=1 | 22.27/0.7660 | 22.29/0.7614 | 22.38/0.7672 | 22.32/0.7617 | **23.04/0.7739** |
| | 5*5/$\sigma_{noise}$=3 | 22.22/0.7403 | 22.15/0.6915 | 22.25/0.7075 | 22.16/0.6932 | **23.35/0.7772** |
| | 5*5/$\sigma_{noise}$=5 | 22.11/0.6964 | 21.85/0.5984 | 21.97/0.6209 | 21.87/0.6010 | **23.27/0.7545** |
| Lenna | 3*3/$\sigma_{noise}$=1 | 29.64/0.8680 | 30.22/0.8763 | 30.03/0.8756 | 30.05/0.8761 | **33.92/0.9016** |
| | 3*3/$\sigma_{noise}$=3 | 29.33/0.8370 | 29.29/0.7922 | 29.23/0.8003 | 29.19/0.7948 | **33.56/0.8930** |
| | 3*3/$\sigma_{noise}$=5 | 28.75/0.7841 | 27.86/0.6765 | 27.93/0.6916 | 27.81/0.6809 | **32.57/0.8612** |
| | 5*5/$\sigma_{noise}$=1 | 28.05/0.8175 | 28.26/0.8142 | 28.15/0.8156 | 28.14/0.8151 | **29.33/0.8334** |
| | 5*5/$\sigma_{noise}$=3 | 27.82/0.7863 | 27.64/0.7299 | 27.60/0.7405 | 27.54/0.7337 | **29.22/0.8255** |
| | 5*5/$\sigma_{noise}$=5 | 27.43/0.7348 | 26.64/0.6180 | 26.70/0.6349 | 26.59/0.6236 | **29.92/0.7990** |
| Boat | 3*3/$\sigma_{noise}$=1 | 27.32/0.7895 | 27.92/0.8143 | 27.76/0.8095 | 27.82/0.8129 | **30.61/0.8490** |
| | 3*3/$\sigma_{noise}$=3 | 27.14/07675 | 27.36/0.7518 | 27.27/0.7546 | 27.26/0.7523 | **30.00/0.8341** |
| | 3*3/$\sigma_{noise}$=5 | 26.78/0.7267 | 26.38/0.6616 | 26.38/0.6705 | 26.32/0.6642 | **29.78/0.8088** |
| | 5*5/$\sigma_{noise}$=1 | 25.70/0.7038 | 25.76/0.7038 | 25.80/0.7064 | 25.74/0.7045 | **26.49/0.7210** |
| | 5*5/$\sigma_{noise}$=3 | 25.57/0.6820 | 25.42/0.6444 | 25.48/0.6540 | 25.42/0.6476 | **26.44/0.7145** |
| | 5*5/$\sigma_{noise}$=5 | 25.33/0.6450 | 24.80/0.5625 | 24.90/0.5765 | 24.82/0.5672 | **26.22/0.6930** |
| Man | 3*3/$\sigma_{noise}$=1 | 27.81/0.8149 | 28.36/0.8399 | 28.31/0.8375 | 28.36/0.8400 | **30.81/0.8673** |
| | 3*3/$\sigma_{noise}$=3 | 27.62/0.7920 | 27.73/0.7750 | 27.75/0.7799 | 27.73/0.7768 | **30.54/0.8585** |
| | 3*3/$\sigma_{noise}$=5 | 27.23/0.7511 | 26.69/0.6830 | 26.79/0.6939 | 26.71/0.6866 | **29.92/0.8265** |
| | 5*5/$\sigma_{noise}$=1 | 26.29/0.7300 | 26.35/0.7306 | 26.37/0.7324 | 26.34/0.7308 | **27.10/0.7448** |
| | 5*5/$\sigma_{noise}$=3 | 26.15/0.7080 | 25.97/0.6702 | 26.02/0.6786 | 25.97/0.6727 | **27.05/0.7397** |
| | 5*5/$\sigma_{noise}$=5 | 25.87/0.6687 | 25.25/0.5820 | 25.36/0.5956 | 25.27/0.5863 | **26.79/0.7148** |
| Couple | 3*3/$\sigma_{noise}$=1 | 27.06/0.7793 | 27.74/0.8130 | 27.56/0.8062 | 27.66/0.8122 | **29.94/0.8455** |
| | 3*3/$\sigma_{noise}$=3 | 26.90/0.7601 | 27.21/0.7575 | 27.09/0.7567 | 27.18/0.7586 | **29.75/08380** |
| | 3*3/$\sigma_{noise}$=5 | 26.57/0.7249 | 26.26/0.6744 | 26.25/0.6801 | 26.24/0.6774 | **29.46/0.8167** |
| | 5*5/$\sigma_{noise}$=1 | 24.51/0.6764 | 25.46/0.6758 | 25.51/0.6803 | 25.41/0.6753 | **26.05/0.6925** |
| | 5*5/$\sigma_{noise}$=3 | 25.30/0.6579 | 25.15/0.6236 | 25.21/0.6332 | 25.13/0.6261 | **26.03/0.6891** |
| | 5*5/$\sigma_{noise}$=5 | 25.07/0.6249 | 24.58/0.5509 | 24.61/0.5609 | 24.58/0.5546 | **25.83/0.6698** |

with the following two metrics: PSNR and SSIM. Table 4 illustrates the PSNR and SSIM values of various SR methods on the six test images shown in Figure 3, with three deviations σ for noise (i.e., $\sigma_{noise}=1$, $\sigma_{noise}=3$, and $\sigma_{noise}=5$) and the Gaussian kernel. Two Gaussian blur kernels were used in the experiments: $3\times3$ Gaussian blur with $\sigma_{blur}=1$, and $5\times5$ Gaussian blur with $\sigma_{blur}=2$. Table 5 presents the PSNR and SSIM values of various SR methods on the six test images in Figure 2, with three deviations σ for noise (i.e., $\sigma_{noise}=1$, $\sigma_{noise}=3$, and $\sigma_{noise}=5$) and the average kernel. In the experiments, the $3\times3$ and $5\times5$ average blur kernels were used.

An analysis of the results presented in Tables 4 and 5 led to the following conclusions: 1) The SISR, SRCNN, and DRRN outperformed the bicubic interpolation method for the six test images and achieved higher average PSNR/SSIM values, which verified the significance of applying the network to the SR problem. 2) the proposed method achieved the best performance on the whole, which demonstrated that the combined data fidelity term was helpful in enhancing the quality of reconstructions.

**4.4 Experiments on simulated data with down-sampling factor 3**

In this subsection, we would show more comparison results with more methods under the case of down-sampling factor equaling to 3. In this sub-section, the noise was also the Gaussian white noise with $\sigma_{noise}=1$. The blur used in this sub-section was Average blur and Gaussian blur, respectively. The Gaussian blur kernels used in this experiment is $3\times3$ Gaussian blur with $\sigma_{blur}=1$. Besides, $5\times5$ average blur kernel is used.

Figure 6 and Figure 7 showed the reconstruction results of different methods for "Couple" image (3×3 Guassian blur with $\sigma_{blur}=1$, Gaussian white noise with $\sigma_{noise}=1$, down-sampling factor 3) and "Man" image (5×5 Average blur, Gaussian white noise with variance 1, down-sampling factor 3). From these two set of images, we can see that the proposed method get the best reconstruction results. In the smooth region, the results of our proposed method had no extra stripes, also less artificial artifacts existed in the edge region.

Table 6 and Table 7 showed the numerical results. PSNR value and SSIM value were used to measure the reconstructed images. The best value of each standard are **Bolded**.

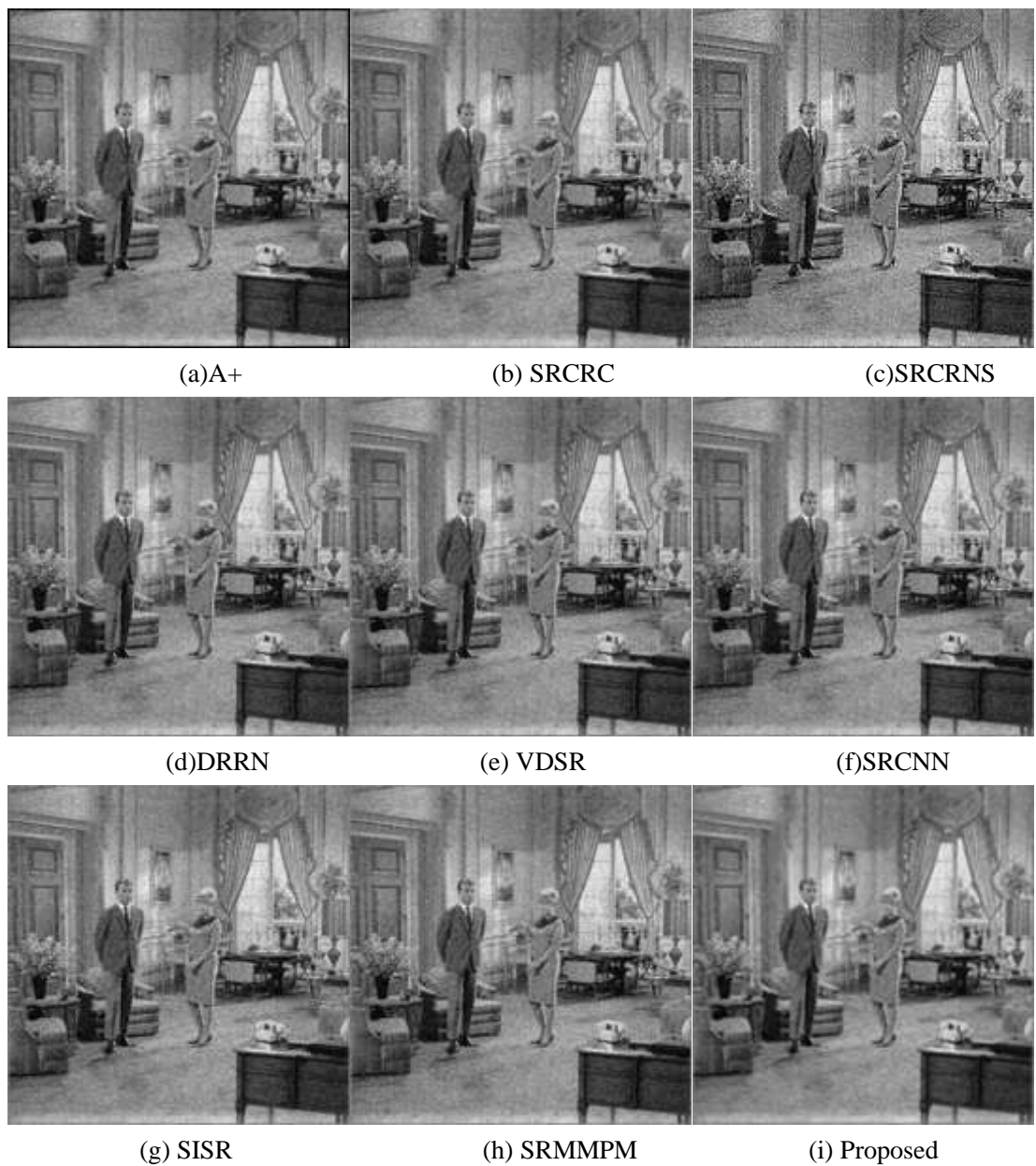

(a)A+            (b) SRCRC            (c)SRCRNS

(d)DRRN            (e) VDSR            (f)SRCNN

(g) SISR            (h) SRMMPM            (i) Proposed

Figure 6. The comparisons of reconstruction results of different methods for "Couple" image under 3×3 Guassian blur with 1, Gaussian white noise with $\sigma_{noise}=1$ down-sampling factor 3.

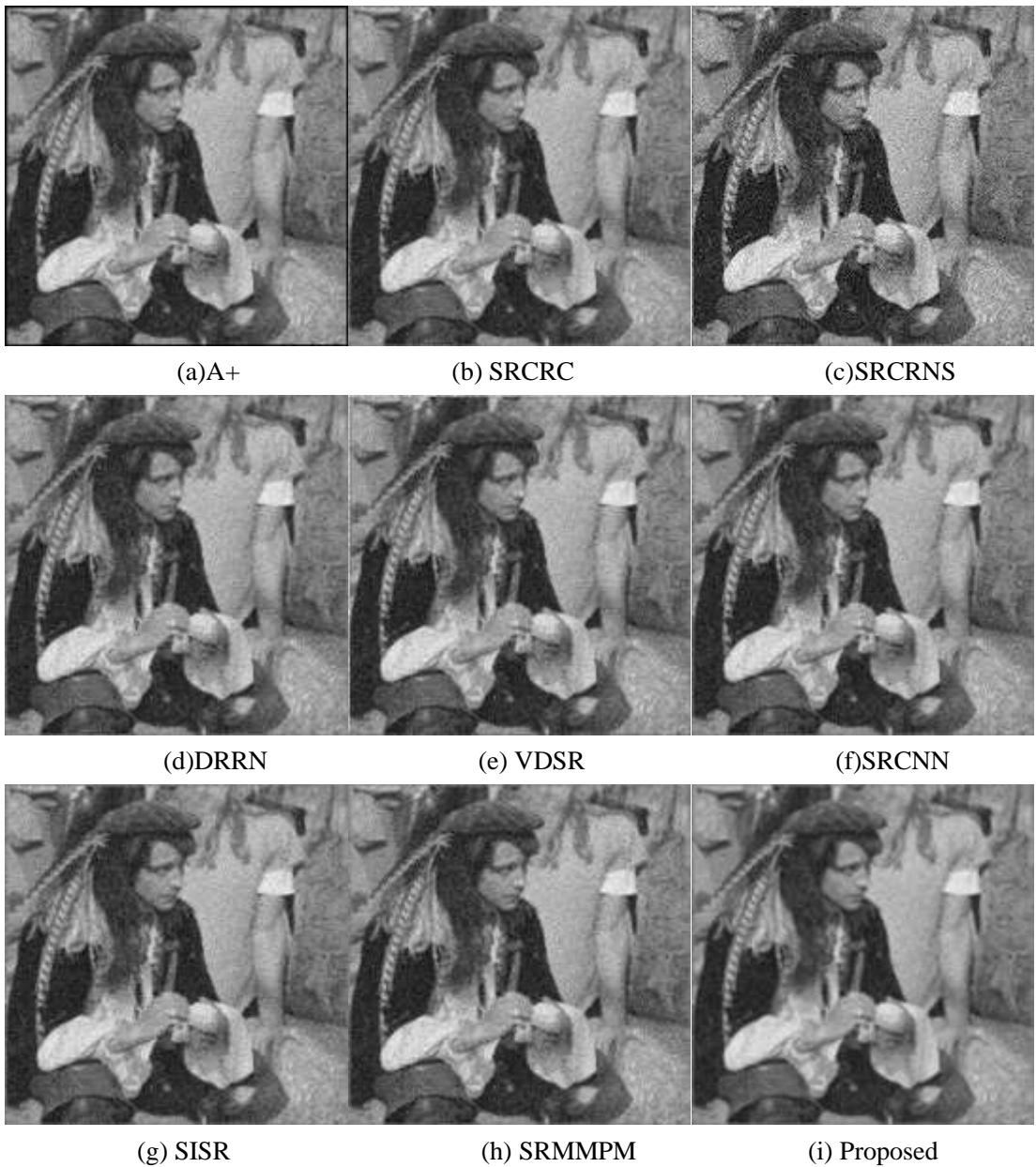

(a)A+          (b) SRCRC          (c)SRCRNS

(d)DRRN          (e) VDSR          (f)SRCNN

(g) SISR          (h) SRMMPM          (i) Proposed

Figure 7. The comparisons of reconstruction results of different methods for "Man" image under 5×5 Average blur, Gaussian white noise with $\sigma_{noise}=1$, down-sampling factor 3.

Table 6 Numerical comparison of different methods under the condition: 3×3 Guassian blur with $\sigma_{blur}=1$, Gaussian white noise with $\sigma_{noise}=1$, down-sampling factor 3.

| Methods | Cameraman | Parrots | Lena | Boat | Man | Couple |
|---|---|---|---|---|---|---|
| A+ | 17.88\0.6084 | 17.49\0.6606 | 21.20\0.6572 | 20.83\0.6342 | 20.94\0.6573 | 20.51\0.6278 |
| SRCRC | 23.17\0.6061 | 23.1\0.6637 | 26.07\0.6473 | 25.11\0.6287 | 25.22\0.6516 | 24.59\0.6236 |
| SRCRNS | 18.19\0.3347 | 17.32\0.3899 | 20.42\0.3605 | 19.84\0.3656 | 19.95\0.3711 | 19.97\0.3746 |
| DRRN | 22.35\0.5259 | 22.15\0.5901 | 26.29\0.6067 | 24.64\0.5522 | 25.08\0.5735 | 24.39\0.5415 |
| VDSR | 24.05\0.6059 | 24.28\0.6682 | 27.54\0.6578 | 26.09\0.6391 | 26.38\0.6572 | 25.49\0.6278 |
| SRCNN | **24.10**\0.6182 | **24.32**\0.6786 | 27.54\0.6618 | 26.07\0.6409 | 26.33\0.6577 | 25.43\0.6256 |
| SISR | 23.77\0.5948 | 23.74\0.6550 | 27.47\0.6419 | 25.83\0.6158 | 26.18\0.6356 | 25.35\0.6064 |
| SRMMPM | 23.92\0.6061 | 24.14\0.6732 | 27.53\0.6575 | 25.97\0.6356 | 26.28\0.6549 | 25.44\0.6234 |
| Proposed | 23.41\**0.7049** | 24.03\**0.7470** | **28.23**\**0.7812** | **26.11**\**0.6947** | **26.56**\**0.7164** | **25.74**\**0.6766** |

Table 7 Numerical comparison of different methods under the condition: 5×5 Average blur, Gaussian white noise with variance 1, down-sampling factor 3.

| Methods | Cameraman | Parrots | Lena | Boat | Man | Couple |
|---|---|---|---|---|---|---|
| A+ | 17.46\0.5379 | 17.01\0.5916 | 20.92\0.6119 | 20.39\0.5568 | 20.56\0.5777 | 20.16\0.5443 |
| SRCRC | 21.67\0.5321 | 21.5\0.5918 | 25.37\0.602 | 23.99\0.5491 | 24.35\0.5696 | 23.75\0.5388 |
| SRCRNS | 19.87\0.3366 | 19.27\0.3874 | 21.93\0.3661 | 21.23\0.352 | 21.47\0.3686 | 21.25\0.3601 |
| DRRN | 22.35\0.5259 | 22.15\0.5901 | 26.29\0.6067 | 24.64\0.5522 | 25.08\0.5735 | 24.39\0.5415 |
| VDSR | 22.36\0.5312 | 22.17\0.5899 | 26.3\0.6059 | 24.66\0.5529 | 25.08\0.5727 | 24.42\0.5424 |
| SRCNN | 22.5\0.5494 | 22.24\0.6048 | 26.33\0.6119 | 24.68\0.5588 | 25.1\0.5772 | 24.41\0.545 |
| SISR | 22.36\0.5399 | 22.09\0.5946 | 26.38\0.5983 | 24.67\0.5499 | 25.14\0.5713 | 24.45\0.5429 |
| SRMMPM | 22.4\0.5384 | 22.18\0.5991 | 26.31\0.6071 | 24.67\0.5547 | 25.08\0.5738 | 24.4\0.5412 |
| Proposed | **22.92**\**0.637** | **22.73**\**0.7026** | **26.81**\**0.722** | **24.78**\**0.6193** | **25.28**\**0.6437** | **24.58**\**0.5999** |

## 4.5 Experiments on real data

In this subsection, the performance of the proposed method and the other compared SR methods on the natural data was discussed. In the first experiment, one LR image ($66 \times 66$) from the "Adyoron" sequence was used.

In the second experiment, one LR image ($96 \times 96$) from the "Alpaca" sequence was used. In the third experiment, one LR image ($96 \times 96$) from the "Foreman" sequence was used. No exact degradation of information was known about these LR frames, and we assumed that in the experiments, the point spread function was a $3 \times 3$ Gaussian kernel with standard deviation equal to 1.

The comparison of different SR approaches for restoring the natural data is shown in Figures 8, 9, and 10. The resolution of Figure 8 was set by a factor 3 and Figure 9 and 10 were set by a factor 2. In the experiments, two different versions of the proposed method were used, namely, the SRBSI and the proposed methods. The compared methods included the bicubic interpolation method, the DRNN method, the SRCNN method, and the SISR method. The visual results showed that the proposed methods were better than the comparison methods. The effect of edge preservation and artifact inhibition can be seen in Figures 8(e) and (i), 9(e) and (i), and 10(e) and (i).

## 5. Conclusion

In this paper, a novel single-frame SR method was proposed. Our model conducts reconstructing and denoising in a unified framework. It obtains the HR image by iterating the improved regularization model and the symmetric residual network. For the regularization model, a new fidelity term was designed to constrain the intensity change and detail (i.e., edge and texture) change simultaneously. Moreover, the reconstruction was improved by the deep network. The proposed SR method has the advantage of making use of the structures of the input and extracting the abundant information from the external data. The experimental results demonstrated that the proposed method outperformed the comparison methods in terms of the visual quality and objective indexes on several test images.

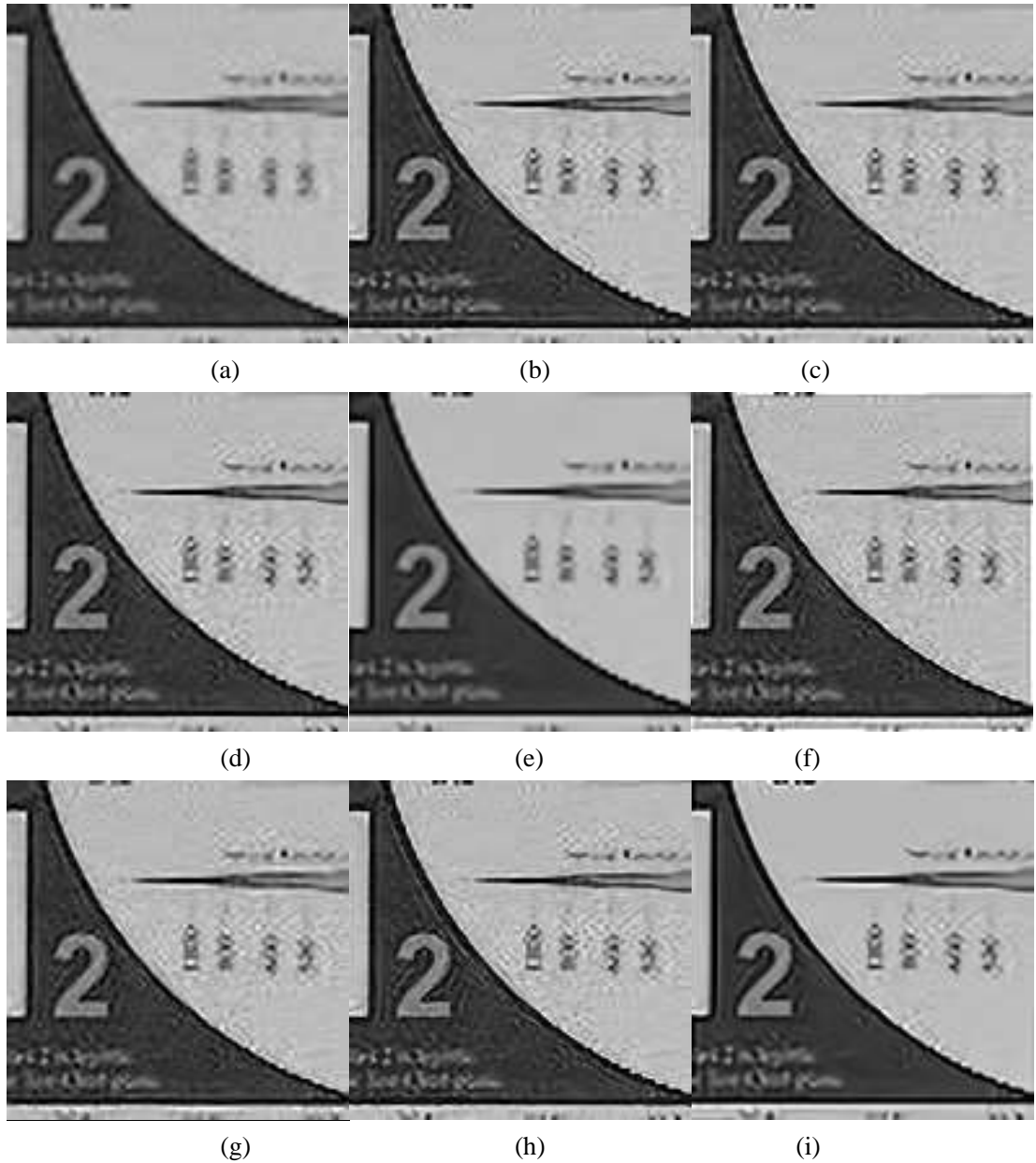

Figure 8. Reconstructed results (×3) of comparison of different methods on the Adyoron sequence. (a) Bicubic; (b) DRNN; (c) SRCNN; (d) SISR; (e) SRBSI; (f) SRCRC; (g)SRMMPM; (h) VDSR; (i) Proposed.

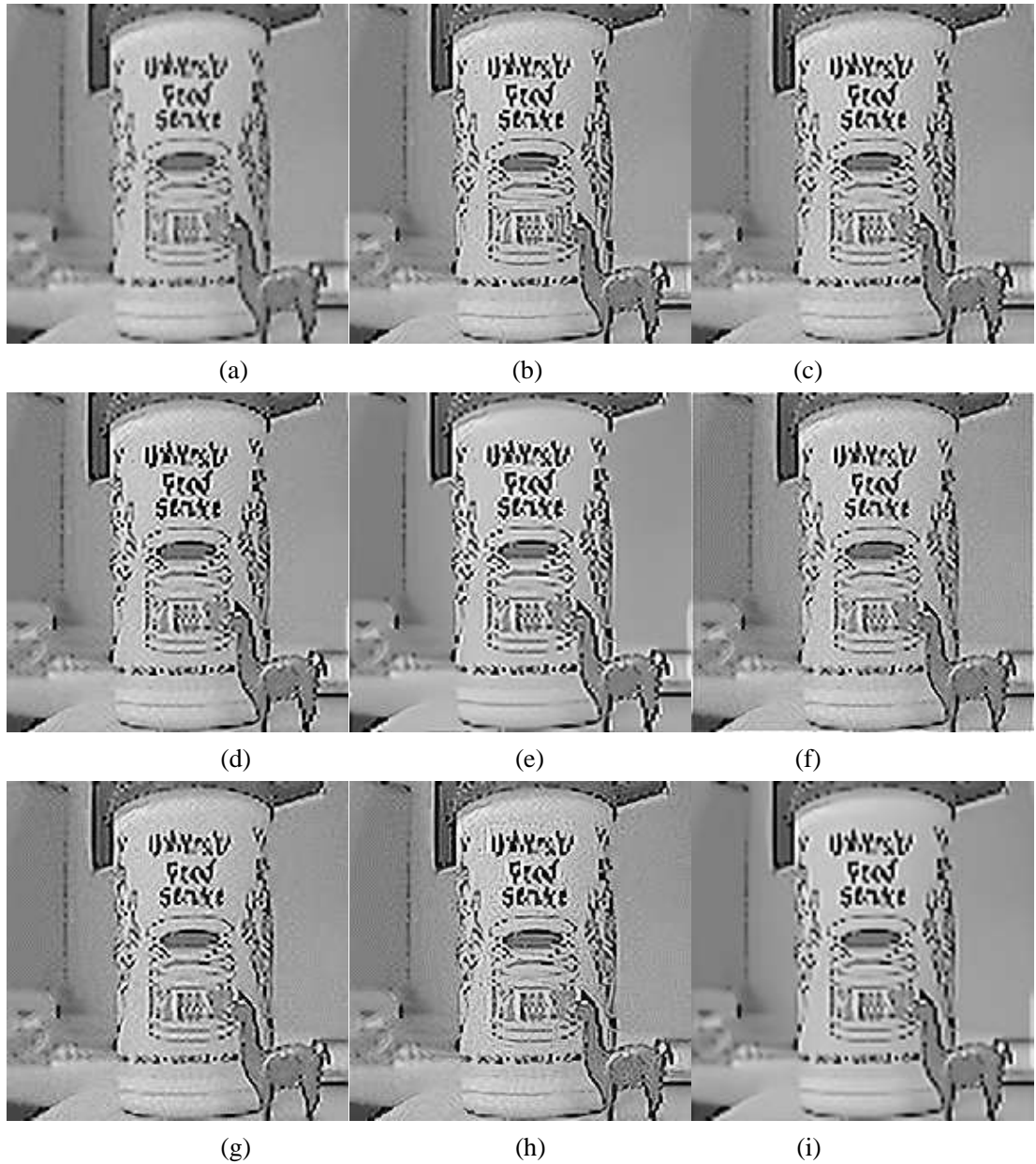

Figure 9. Reconstructed results (×2) of comparison of different methods on the Alpaca sequence. (a) Bicubic; (b) DRNN; (c) SRCNN; (d) SISR; (e) SRBSI; (f) SRCRC; (g)SRMMPM; (h) VDSR; (i) Proposed.

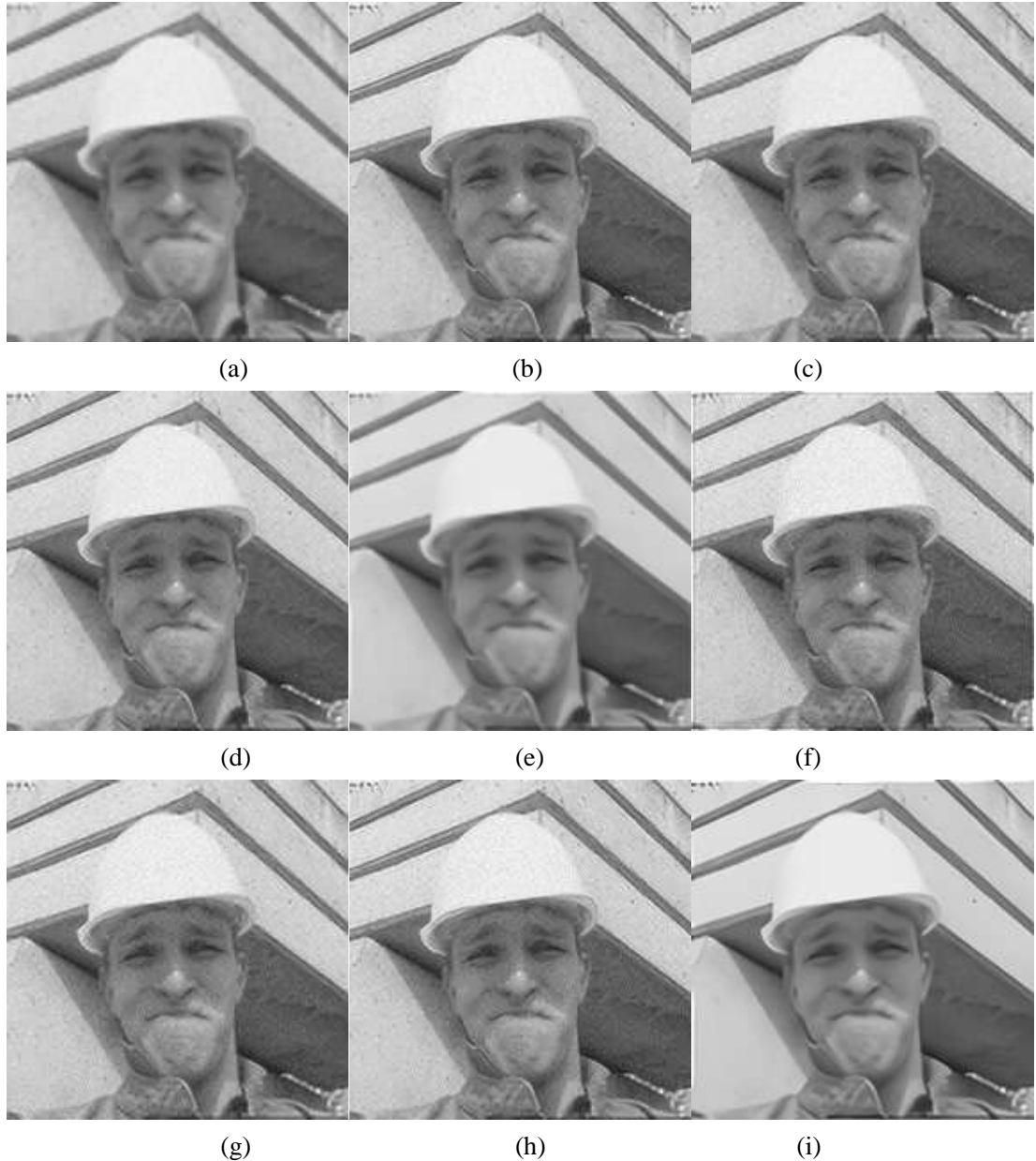

Figure 10. Reconstructed results (×2) of comparison of different methods on the Foreman sequence. (a) Bicubic; (b) DRNN; (c) SRCNN; (d) SISR; (e) SRBSI; (f) SRCRC; (g)SRMMPM; (h) VDSR; (i) Proposed.

## Acknowledgement

This work is supported by National Natural and Science Found of China (No.61802213) Shandong Provincial Natural Science Found (No.ZR2017LF016, ZR2018LF004).

## Reference

[1] A. Lucas, M. Iliadis, R. Molina, A. Katsaggelos, Using Deep Neural Networks for Inverse Problems in Imaging: Beyond Analytical Methods. IEEE Signal Processing

Magazine, 35 (1): 20-36, 2018

[2] F. Zhou, X. Li, Z. Li, High-Frequency Details Enhancing DenseNet for Super-Resolution. Neurocomputing, 290: 34-42, 2018

[3] C. Zou, Y. Xia, Bayesian dictionary learning for hyperspectral image super resolution in mixed Poisson–Gaussian noise. Signal Processing: Image Communication, 60: 29-41, 2018

[4] W. Shi, J. Caballero, F. Huszar, J. Totz, A. P. Aitken, R. Bishop, D. Rueckert, Z. Wang, Real-Time Single Image and Video Super-Resolution Using an Efficient Sub-Pixel Convolutional Neural Network. IEEE Conference on Computer Vision and Pattern Recognition, 1874-1883, Las Vegas, NV, United States, 2016

[5] Y. Zhang, Q. Fan, F. Bao, Y. Liu, C. Zhang, Single-Image Super-Resolution Based on Rational Fractal Interpolation. IEEE Transactions on Image Processing, 27(8): 3782-797, 2018

[6] K. Konstantoudakis, L. Vrysis, N. Tsipas, C. Dimoulas, Block unshifting high-accuracy motion estimation: A new method adapted to super-resolution enhancement. Signal Processing: Image Communication, 65: 81-93, 2018

[7] I. Mourabit, M. Rhabi, A. Hakim, A. Laghrib, E. Moreau, A new denoising model for multi-frame super-resolution image reconstruction. Signal Processing, 132 (C): 51-65, 2017

[8] S. Huang, J. Sun, Y. Yang, Y. Fang, P. Lin, Y. Que, Robust Single-Image Super-Resolution Based on Adaptive Edge-Preserving Smoothing Regularization. IEEE Transactions on Image Processing, 27(6): 2650-2663, 2018

[9] A. Laghrib, M. Ezzaki, M. Rhabi, A. Hakim, P. Monasse, S. Raghay, Simultaneous deconvolution and denoising using a second order variational approach applied to image super resolution. Computer Vision & Image Understanding, 168: 50-63, 201

[10] M.L. Zhang, C. Desrosiers, High-quality Image Restoration Using Low-Rank Patch Regularization and Global Structure Sparsity. IEEE Transactions on Image Processing, 28(2): 868-879, 2019

[11] K.B. Zhang, Z. Wang, J. Li, X.B. Gao, Z.G. Xiong, Learning recurrent residual regressors for single image super-resolution. Signal Processing, 154: 324-337, 2019

[12] Y X. Yang, H.Y. Mei, J.Q. Zhang, K. Xu, B.C. Yin, Q. Zhang, X.P. Wei, DRFN: Deep Recurrent Fusion Network for Single-Image Super-Resolution with Large Factors. IEEE Transactions on Multimedia, 21(2): 328-337, 2019

[13] Z.J. Wang, B. Chen, H. Zhang, H.W. Liu, Variational probabilistic generative framework for single image super-resolution. Signal Processing, 156: 92-105, 2019

[14] H. Chen, X. He, L. Qing, Q. Teng, Single Image Super-Resolution via Adaptive Transform-Based Nonlocal Self-Similarity Modeling and Learning-Based Gradient Regularization. IEEE Transactions on Multimedia, 19 (8): 1702-1717, 2017

[15] Y. Zhang, F. Shi, J. Cheng, L. Wang, P. Yap, D. Shen, Longitudinally Guided Super-Resolution of Neonatal Brain Magnetic Resonance Images. IEEE Transactions on Cybernetics, 2018, (Early Access)

[16] A. Kyrillidis, Simple and practical algorithms for lp-norm low-rank approximation. arXiv:1805.09464v1:16pages, 2018

[17] M. Chen, C. Tang, J. Zhang, Z. Lei, Image decomposition and denoising based

on Shearlet and nonlocal data fidelity term. Signal Image & Video Processing, 2018 (7) :1-8, 2018

[18] N. Darginis, R. Achanta, S. Susstrunk, Single image reflection suppression. In: IEEE Conference on Computer Vision and Pattern Recognition (CVPR). 1752-1760, 2017

[19] X. Yang, J. Zhang, Y. Liu, X. Zheng, K. Liu, Super-resolution image reconstruction using fractional-order total variation and adaptive regularization parameters. Visual Computer, 2018 (In Press, doi.org/10.1007/s0037)

[20] Z. Jiang, Y. Hou, H. Yue, J. Yang, C. Hou, Depth Super-resolution from RGB-D Pairs with Transform and Spatial Domain Regularization. IEEE Transactions on Image Processing, 27(5): 2587-2602, 2018

[21] J. Liu, W. Yang, X. Zhang, Z. Guo, Retrieval compensated group structured sparsity for image super-resolution. IEEE Transactions on Multimedia, 19(2):302-316, 2017

[22] J. Jiang, X. Ma, C. Chen, T. Lu, Z. Wang, and J. Ma, Single image super-resolution via locally regularized anchored neighborhood regression and nonlocal means. IEEE Transactions on Multimedia, 19(1): 15-26, 2017

[23] S. Zhao, H. Liang, M. Sarem, A Generalized Detail-Preserving Super-Resolution method. Signal Processing, 120:156-173, 2016

[24] K. Zhang, J. Li, H. Wang, X. Liu, X. Gao, Learning local dictionaries and similarity structures for single image super-resolution. Signal Processing, 142: 231-243, 2018

[25] Q. Yan, Y. Xu, X. Yang, T. Nguyen, Single Image Super-Resolution Based on Gradient Profile Sharpness. IEEE Transactions on Image Processing, 24(10): 3187-3202, 2015

[26] K. Chang, P. Ding, B. Li, Single Image Super Resolution Using Joint Regularization. IEEE Signal Processing Letters, 25(4): 596 -600, 2018

[27] C. Dong, C.C. Loy, K. He, X. Tang, Image super-resolution using deep convolutional networks. IEEE Transactions on Pattern Analysis and Machine Intelligence, 38(2): 295-307, 2016

[28] J. Kim, J. Kwon Lee, K. Mu Lee, Accurate image super-resolution using very deep convolutional networks. in: Proceedings of the IEEE Conference on Computer Vision and Pattern Recognition, 1646–1654, 2016

[29] K. He, X. Zhang, S. Ren, J. Sun, Deep residual learning for image recognition. in: Proceedings of the IEEE Conference on Computer Vision and Pattern Recognition, 770–778, 2016

[30] Y. Tai, J. Yang, X. Liu, Image super-resolution via deep recursive residual network. in: Proceedings of the IEEE Conference on Computer Vision and Pattern Recognition, 3147–3155, 2017

[31] W.-S. Lai, J.-B. Huang, N. Ahuja, M.-H. Yang, Deep laplacian pyramid networks for fast and accurate super-resolution. in: Proceedings of the IEEE Conference on Computer Vision and Pattern Recognition, 624–632, 2017

[32] J. Lu, W.D. Hu, Y. Sun, A Deep Learning Method for Image Super-Resolution Based on Geometric Similarity. Signal Processing: Image Communication, 70:

210-219, 2019

[33] W.H. Yang, J.S. Feng, J.C.Yang, F. Zhao, J.Y. Liu, Z.M. Guo, S.C. Yan, Deep Edge Guided Recurrent Residual Learning for Image Super-Resolution. IEEE Transactions on Image Processing, 26(12): 5895-5907, 2017

[34] M. Zareapoor, M.E. Celebi, J.Yang, Diverse adversarial network for image super-resolution. Signal Processing: Image Communication, 74: 191-200, 2019

[35] Z.D. Zhang, X.R. Wang, C.K. Jung, DCSR: Dilated convolutions for single image super-resolution. IEEE Transactions on Image Processing, 28(4): 1625-1635, 2019

[36] Y. Zheng, X. Cao, Y. Xiao, X.Y. Zhu, J. Yuan, Joint residual pyramid for joint image super-resolution. Journal of Visual Communication and Image Representation, 58: 53-62, 2019

[37] Y.X. Chen, H.D.Tan, L.M.Zhang, J.Zhou, Q. Lu, Hybrid image super-resolution using perceptual similarity from pre-trained network. Journal of Visual Communication and Image Representation, 60: 229-235, 2019

[38] K. Zhang, W. Zuo, S. Gu, L. Zhang, Learning Deep CNN Denoiser Prior for Image Restoration. IEEE Conference on Computer Vision and Pattern Recognition (CVPR), 2808 – 2817, 2017

[39] J. Zhang, J. Pan, W. Lai, R. Lau, M. Yang, Learning Fully Convolutional Networks for Iterative Non-blind Deconvolution. in: proceedings of the IEEE Computer Society Conference on Computer Vision and Pattern Recognition, 6969 -6977, 2017

[40] V. Lempitsky, A.Vedaldi, D. Ulyanov, Deep Image Prior. in: proceedings of the IEEE Computer Society Conference on Computer Vision and Pattern Recognition, 9446-9454, 2018

[41] R. Timofte, V. De Smet, and L. Van Gool, A+: Adjusted anchored neighborhood regression for fast super-resolution. Lecture Notes in Computer Science, 9006: 111-126, 2015

[42] Y.B. Zhang, Y.L. Zhang, J. Zhang, D. Xu, Y. Fu, Y.S. Wang, X.Y. Ji, Q.H. Dai, Collaborative Representation Cascade for Single-Image Super-Resolution. IEEE Transactions on Systems, Man, and Cybernetics: Systems, 49(5): 845-860, 2019

[43] K. Chang, P. Ding, B.X. Li, Single image super-resolution using collaborative representation and non-local self-similarity. Signal Processing, 149: 49-61, 2018

[44] Y.F. Huang, J. Li, X.B. Gao, L.H. He, W. Lu, Single Image Super-Resolution via Multiple Mixture Prior Models. IEEE Transactions on Image Processing, 27(12): 5904-5917, 2018